\documentclass[12pt]{iopart}
\usepackage{graphicx}
\usepackage{bbold}
\usepackage{cite}

\begin{document}

\title{Electronic correlations in vanadium chalcogenides: BaVSe$_3$
versus BaVS$_3$}

\author{Daniel Grieger, Lewin Boehnke and Frank Lechermann}
\address{I. Institut f\"ur Theoretische Physik, Universit\"at~Hamburg, 
Jungiusstr.~9, D-20355~Hamburg, Germany}
\ead{daniel.grieger@physnet.uni-hamburg.de}

\begin{abstract}
Albeit structurally and electronically very similar, at low temperature
the quasi-one-dimensional vanadium sulfide BaVS$_3$ shows a metal-to-insulator 
transition via the appearance of a charge-density-wave state, while BaVSe$_3$ 
apparently remains metallic down to zero temperature. This different 
behavior upon cooling is studied by means of density functional theory and 
its combination with the dynamical mean-field theory and the 
rotationally-invariant slave-boson method. We reveal several subtle 
differences between these chalcogenides that provide indications for the 
deviant behavior of BaVSe$_3$ at low temperature. In this regard, a smaller 
Hubbard $U$ in line with an increased relevance of the Hund's exchange $J$ 
plays a vital role.
\end{abstract}

\pacs{71.45.Gm, 71.45.Lr, 71.30.+h, 73.20.Mf, 71.15.Mb}
\submitto{\JPCM}
\noindent(Some figures in this article are in color only in the electronic version)

\section{Introduction}

The discussion of electronic correlation effects in
$3d$-transition-metal compounds appears often dominantly reserved for
the physics of oxides, e.g.  for cuprates, manganites, ruthenates, or
more recently cobaltates. However there is increasing awareness that
strong correlation effects are also relevant for many intriguing
electronic phase transitions of various chalcogenides such as the
metal-insulator transition (MIT) in NiS$_{2-x}$Se$_x$~\cite{oga79},
the charge-density wave (CDW) transition in NbSe$_3$~\cite{mon76} or
the recent discovery of higher-temperature superconductivity in
Fe-chalcogenides.  In this respect, the vanadium chalcogenides with
the quasi-two-dimensional X-VS$_2$ misfit-layer structures~\cite{nis96}
and the quasi-one-dimensional (1D) BaV(S,Se)$_3$ systems offer a unique
opportunity to study the interplay of correlation effects and
effective low-dimensionality in these transition-metal compound systems.

Here we want to concentrate on the subtle comparison of the quasi-1D
barium vanadium chalcogenide systems. It will become evident that the
vanadium selenide BaVSe$_3$ is in effect rather similar to the
corresponding vanadium sulfide BaVS$_3$ in many respects. The latter
compound has been the subject of various studies concerning its
complex electronic structure throughout the last
decades~\cite{mas79,mat91,nak94,gra95,boo99,for00,wha03,fag03,lec07}.
This sulfide exhibits three continuous phase transitions with
decreasing temperature, one of which is a highlighting
metal-to-paramagnetic-insulator transition at
$T_{\mathrm{MIT}}~\sim~$70K. Though the exact details of the scenario
are still to be revealed, strong evidence exists~\cite{ina02,fag03}
that this transition is associated with a CDW
instability. Interestingly, such a transition cannot be found in the
similar selenide compound, which remains metallic down to very low
temperatures~\cite{akr08}. This raises questions about the delicate
electronic and structural differences that anticipate a possible CDW
in these chalcogenides. Hence the present investigation is devoted to
a detailed comparison between the correlated electronic structure of
BaVS$_3$ and BaVSe$_3$.

Concerning the structural properties, many features of BaVSe$_3$ are
akin to the ones of BaVS$_3$. At room temperature, both materials
exhibit a hexagonal structure in which the vanadium atoms form
straight chains in the crystallographic $c$~direction. The intrachain
V-V distance is less than half of the interchain distance, so that one
is tempted to identify a quasi-1D substructure in these
compounds. The chains become slightly distorted in the $ab$ plane with
a zigzag distortion in the $bc$ plane at the continuous transition
temperature $T_S~\sim~$(240K, 290K) (sulfide,
selenide)~\cite{say82,kel79}. Thus, below this temperature regime, an
orthorhombic crystal structure is stabilized for both
materials. Furthermore, the additional metal-to-insulator transition
of the sulfide is accompanied by a doubling of the unit cell and a
transition to the monoclinic crystal system~\cite{ina02}, whereas the
selenide remains orthorhombic down to very low temperature. Both
chalcogenides display local-moment behavior at ambient $T$ and a final
continuous magnetic ordering transition. While in BaVS$_3$ an
incommensurate antiferromagnetic (N\'eel) order develops below 
$T_{\rm N}~\sim~$30K~\cite{nak00}, BaVSe$_3$ undergoes a ferromagnetic
transition at $T_{\rm C}~\sim~$43K~\cite{kel79,yam01}.

The electronic structure of both materials in the low-energy regime is
dominated by the nominal $\mathrm{V}^{4+}$ valence, so that they can
be described as $3d^1$ compounds with a $t_{2g}$-like manifold at the Fermi 
level~\cite{mat91,wha03,lec07}. The latter consists of an $A_{1g}$-like 
orbital pointing along the vanadium chain direction as well as two $E_g$-like 
orbitals pointing inbetween the neighboring sulfur/selenium ions, 
respectively. Due to strong hybridization with S($3p$)/Se($4p$), the V($e_g$) 
states have mainly high-energy weight and may be integrated out in a first 
low-energy approximation. Hence the essential physics of these systems can be 
described within a three-band model approach~\cite{lec07}. The two $E_g$ 
states are degenerate in the hexagonal structure, but split into an $E_{g1}$ 
and an $E_{g2}$ state in the orthorhombic regime. Due to their somewhat
isolated orbital space, the hybridization with the environment is comparably 
small. For this reason, they form very narrow bands that show almost no 
dispersion. In contrast, the $A_{1g}$ orbitals display a larger dispersion 
along the V chain direction. Hence, in a minimal model, the physics of these 
materials essentially boils down to the interplay of an itinerant and two 
localized states sharing a single electron.

\section{Theoretical Framework}

Due to the obviously delicate electronic systems in BaVS$_3$ and BaVSe$_3$,
with narrow bands at the Fermi level and local-moment physics, methods are 
needed that allow us to take into account many-body effects explicitly, i.e.
beyond the standard density functional theory (DFT) in its local density
approximation (LDA) to the exchange-correlation energy. In the
present work, the opportunity is taken to compare two different approaches
for this task, namely the LDA+DMFT~\cite{ani97,lic98} framework in conjunction
with a highly-evolved quantum-Monte-Carlo impurity solver as well as the 
combination of LDA with the rotationally-invariant slave-boson 
formalism (RISB)~\cite{li89,lecgeo07}. 

The DFT parts of the following calculations have been performed using a 
mixed-basis pseudopotential (MBPP) code~\cite{mbpp_code}, which utilizes 
norm-conserving pseudopotentials and a combined basis set of plane waves as 
well as non-overlapping localized functions. From the LDA band structure, by 
means of the maximally-localized Wannier function 
construction~\cite{mar97,sou01,mos08}, a low-energy Kohn-Sham (KS) Hamiltonian 
$H_{\rm KS}^{(\mathcal C)}(\mathbf k)$ within a correlated subspace ${\cal C}$ is 
extracted~\cite{lec06}. In this work we restrict all self-energy effects 
to this subspace and hence may write the local interacting Green's function as
\begin{equation}
G_{mm'} (i \omega_n) = \sum \limits_{\mathbf k} \left\{\left [ (i
  \omega_n + \mu) \mathbb 1 - 
H_{\rm KS}^{(\mathcal C)} (\mathbf k) -
  \mathbf \Sigma^{(\mathcal C)} (i \omega_n) \right ]^{-1}\right\}_{mm'}
\quad.\label{greenf}
\end{equation}
Here $\omega_n$ are Matsubara frequencies and
$\mathbf{\Sigma}^{(\mathcal C)}$ is the self-energy matrix for the
correlated orbitals denoted by $m,m'$. Note that (\ref{greenf}) is
written in the DMFT approximation, i.e., with a purely local
self-energy (for a review see e.g.~\cite{geo96}). The mean-field
version of RISB which is put into practice in the present work also
has a local self-energy by construction. Moreover, the RISB
self-energy only captures the linear term of the frequency dependence,
expressed via the quasiparticle weight $Z$, as well as a local static
part~\cite{lec07}.  Thus without going into details, the main
differences between DMFT and RISB (in the mean-field version) is the
neglection of the frequency dependence of the higher-energy
excitations in the latter framework. However, as a saddle-point
approximation the performance is fairly efficient, with reliable
qualitative (and often even good quantitative) results in most cases.
It is also important to realize that although we use here a
lattice-implementation of the RISB method, this technique may equally
well be utilized as an impurity solver for standard DMFT. Due to the
inherent local nature at the saddle-point, this specific DMFT impurity
solution is then identical to the direct lattice-calculation result.

In the present work, the explicit quantum impurity problem within DMFT is 
solved using the continuous-time quantum-Monte-Carlo (CTQMC) approach 
employing the hybridization-expansion method~\cite{wer06}. For the actual 
computations the recent implementation by Ferrero and Parcollet
(see \cite{fer09,kotliar_review}) is employed. Both 
frameworks, i.e., DMFT~(CTQMC) and RISB, allow for non-density-density terms in 
the many-body Hamiltonian, which we restrict in the present case to spin-flip 
and pair-hopping processes. Thus the following interacting Hamiltonian for the
minimal modeling may be used:
\begin{eqnarray}
\label{Eqn:INTERACTINGHAMILTONIAN}
\hat H_{\mathrm{int}} &=& U\sum_m\hat n_{m \uparrow} \hat
n_{m \downarrow}+ \frac 12 \sum \limits _{m \ne m'} \sum \limits _\sigma
\left [ {U'} \, \hat n_{m \sigma} \hat n_{m' \bar \sigma} + (U' - J) \,
\hat n_{m \sigma}\hat n_{m' \sigma} \right ] \nonumber \\
&+& \frac {1} 2 \sum \limits _{m \ne m'}\sum \limits _\sigma \left 
[J \, \hat d^\dagger_{m \sigma} \hat d^\dagger_{m' \bar
  \sigma} \hat d^{\hfill}_{m \bar \sigma} \hat d^{\hfill}_{m' \sigma}
+ J_{\mathrm C} \,\hat d^\dagger_{m \sigma} \hat d^\dagger_{m \bar \sigma}
 \hat d^{\hfill}_{m' \bar \sigma} \hat d^{\hfill}_{m' \sigma} \right ]\quad,
\end{eqnarray}
where $\hat n_{m\sigma}$~=~$\hat d^\dagger_{m \sigma} \hat
d^{\hfill}_{m \sigma}$, with $\hat d^{(\dagger)}_{m\sigma}$ as the
electron annihilation (creation) operator for orbital $m$ and spin
$\sigma$.  Throughout the calculations, the choice $U'$~=~$U$-$2J$ and
$J_{\mathrm C}$~=~$J$, appropriate for $t_{2g}$ systems, is made.

\section{Results}
\subsection{LDA studies}

The following LDA comparison between BaVS$_3$ and BaVSe$_3$ builds up on
the orthorhombic crystal structure (space group $Cmc2_1$), since we
are mainly interested in elucidating the possibility of the CDW state
in both systems. Note that the CDW instability is not only absent in
the selenide compound, but also vanishes in the sulfide at high
pressure~\cite{for00}. This raised the suscpicion that the selenide
compound may be interpreted as the high-pressure analog of the
sulfide compound~\cite{akr08}. In any case, theoretically
investigating BaVSe$_3$ should surely also provide more information
about the origin of the apparent CDW instability in BaVS$_3$.

The existence of the orthorhombic phase of the selenide has been
reported already on the occasion of its first synthesis by Kelber {\sl
  et al.}~\cite{kel79}, but to the present authors' knowledge no
detailed experimental structural data is available in the
literature. For this reason, the atomic positions used for
$Cmc2_1$-BaVSe$_3$ are revealed by structural relaxation within the
LDA using the MBPP code, while the crystal parameters are based on
similar recent calculations performed by Akrap {\sl et
  al.}~\cite{akr08}. In contrast, structural data obtained by neutron
diffraction by Ghedira {\sl et~al.}~\cite{ghe86} are used for
BaVS$_3$. Note that an LDA structural relaxation of this experimental
data for $Cmc2_1$-BaVS$_3$ does result in only marginal changes of the
atomic positions, therefore an influence thereof on the following
studies cannot be found. The crystal parameters $(a,b,c)$ for one unit
cell consisting of two formula units of BaVS$_3$ or BaVSe$_3$,
respectively, in au read (12.77, 21.71, 10.58) for BaVS$_3$ and
(13.34, 22.68, 11.06) for BaVSe$_3$. For both materials the ideal
quasi-1D chains of vanadium ions are slightly distorted in the
orthorhombic phase, leading to zig-zag chains. This effect turns out
to be slightly larger in the selenide (relative displacement: 0.025)
than in the sulfide (0.021).
\begin{figure}[h]
\centering
\includegraphics[width=6.5cm,clip]{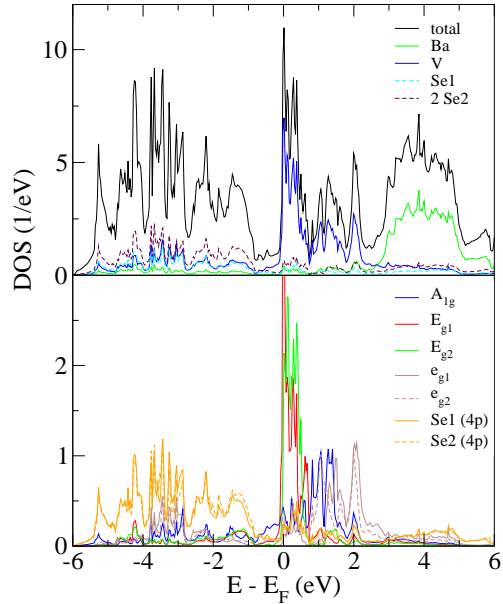}
\caption{\label{Img:LDADOS_BAVSE3}
  LDA density of states of $\mathrm{BaVSe_3}$.}
\end{figure}
\begin{figure}[h]
\centering
\includegraphics[width=6.5cm,clip]{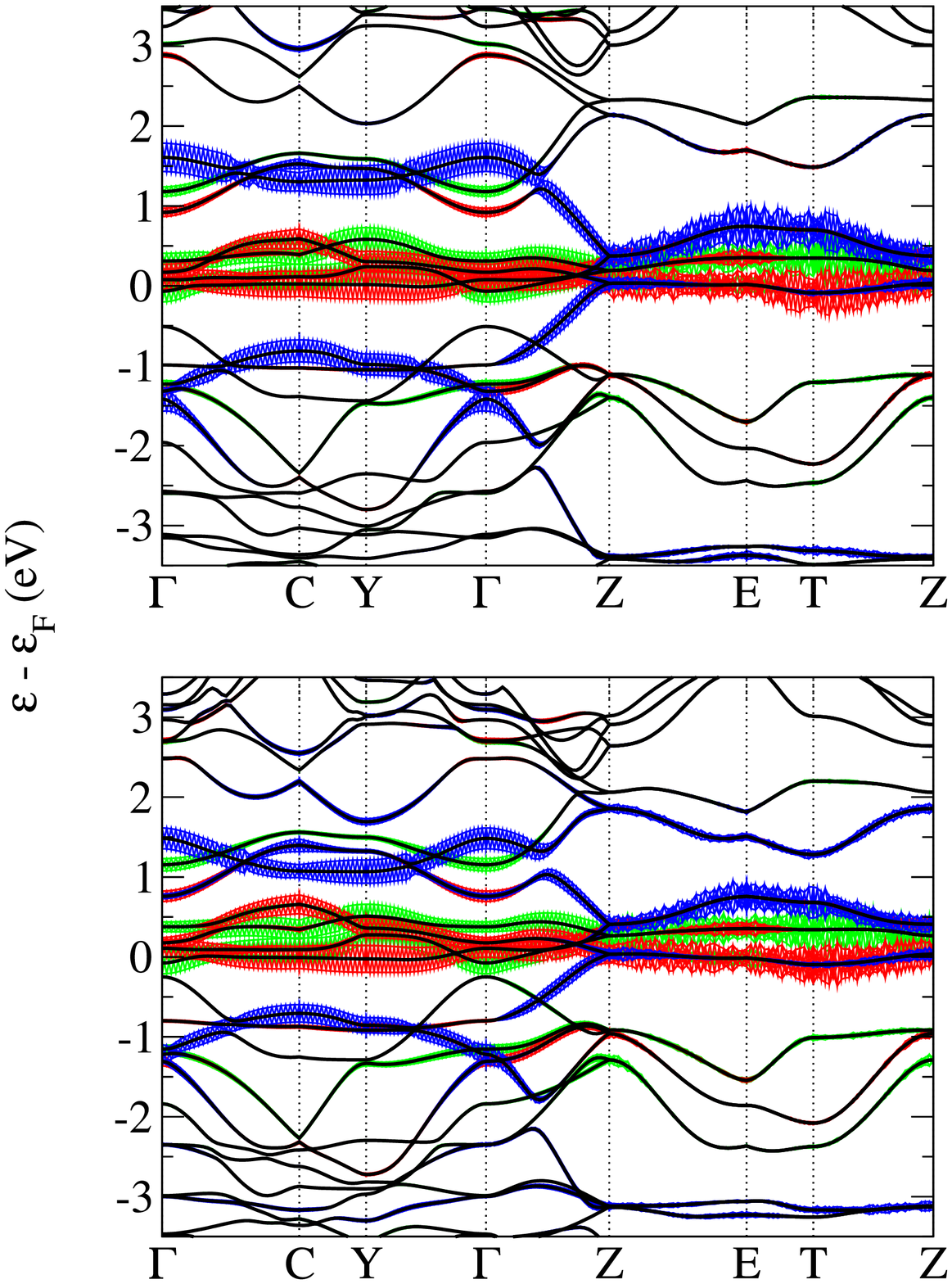}
\hspace*{0.25cm}
\includegraphics[width=4.5cm,clip]{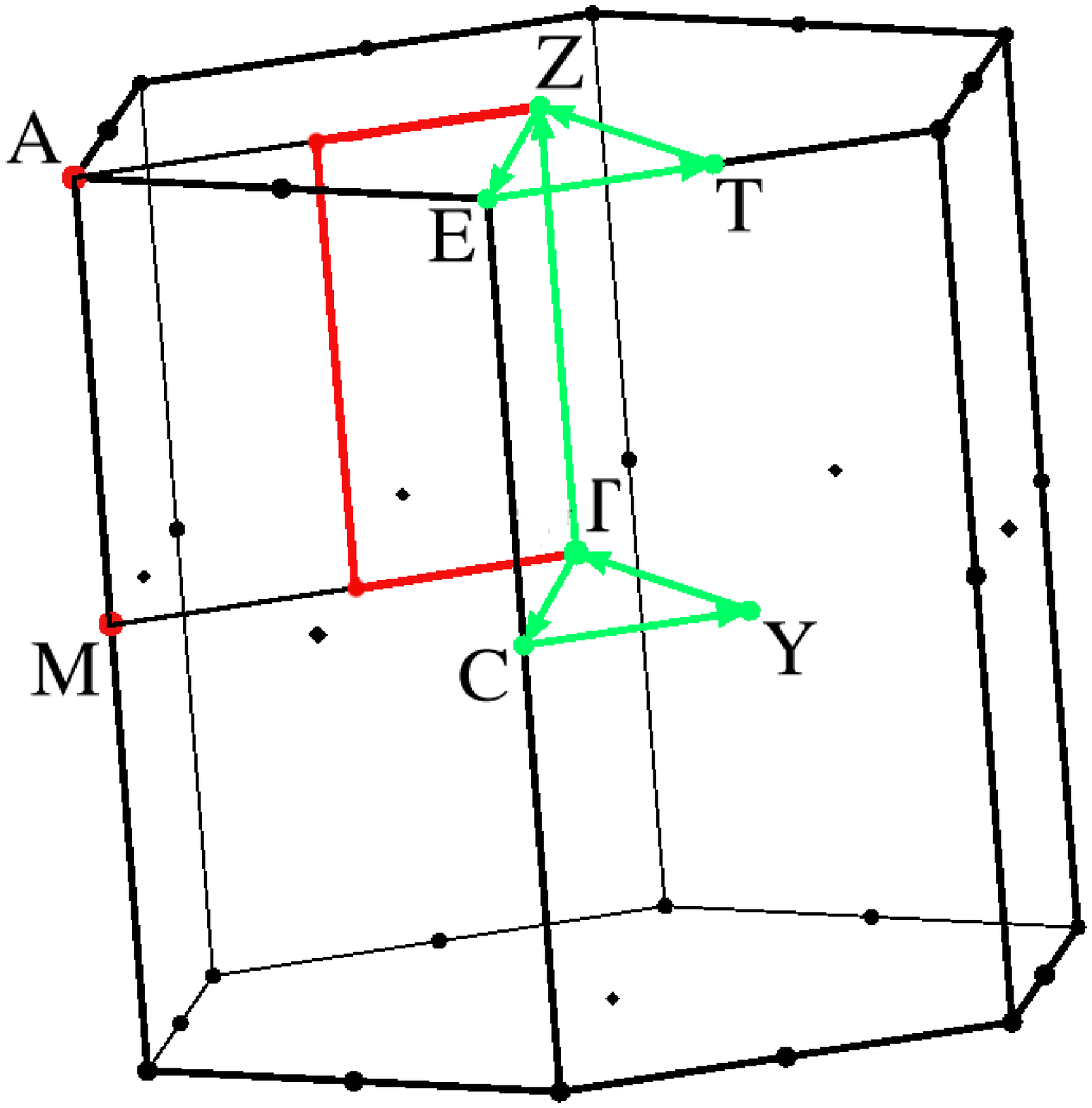}
\caption{\label{Img:BDSTRUC_LDA_FATBANDS}
  Left: LDA band structure of $\mathrm{BaVS_3}$~(top) and
  $\mathrm{BaVSe_3}$~(bottom). The fatband broadening depicts the weight
  of $t_{2g}$ orbitals on the bands. Color coding: $A_{1g}$, blue/dark,
  $E_{g1}$, red/medium and $E_{g2}$, green/light. Right: Brillouin zone of the
  orthorhombic structure with relevant points/directions.}
\end{figure}
Concerning the electronic structure, figure~\ref{Img:LDADOS_BAVSE3}
shows the LDA density of states (DOS) of BaVSe$_3$ in a comparatively
large energy window around the Fermi level. Analogous to
BaVS$_3$~\cite{lec07}, it can be seen that the low-energy physics of the
material may be dominantly described by the abovementioned $t_{2g}$
manifold. Thus a downfolding to an effective three-band model
consisting of a broader $A_{1g}$-like level and narrower $E_{g1}$-like
and $E_{g2}$-like levels is a suitable approximation for the essential
physics.  The accordance with the LDA DOS of the sulfide (shown in
\cite{lec07}) on this level of comparison is rather striking and only
minor differences show up.  Figure~\ref{Img:BDSTRUC_LDA_FATBANDS}
compares the band structure of BaVSe$_3$ and of BaVS$_3$ in a smaller
energy window, with 'fatbands' showing the weight of the
symmetry-adapted $t_{2g}$ states on the specific bands. Again both
band structures are similar, yet subtle difference may be
observed. First the Se($4p$) band at the $\Gamma$ point is much
closer to the Fermi level than the corresponding S($3p$) band in
BaVS$_3$, which may lead to a smaller Hubbard $U$ for the minimal 3-band
model due to increased screening.  Furthermore the bandwidth
difference between the three $t_{2g}$ bands is reduced, i.e., the
$A_{1g}$-like bands become narrower and the $E_g$-like ones
broader. Hence the total $t_{2g}$ bandwidth in BaVSe$_3$ is slightly smaller
($\sim$2.4 eV) than in BaVS$_3$ ($\sim$2.7 eV). This may be explained by
the fact that since the larger Se ions give rise to increased
interatomic distances, a smaller $A_{1g}$-like hopping amplitude along
the chains results.  As a general feature, the overall $A_{1g}$
contribution to the bands seems somewhat more distributed in
BaVSe$_3$. This enhanced spread in energy is again providing hints
towards a larger degree of, eventually more isotropic, localization in
the selenide.

\begin{figure}[h]
\centering
\includegraphics[width=3cm]{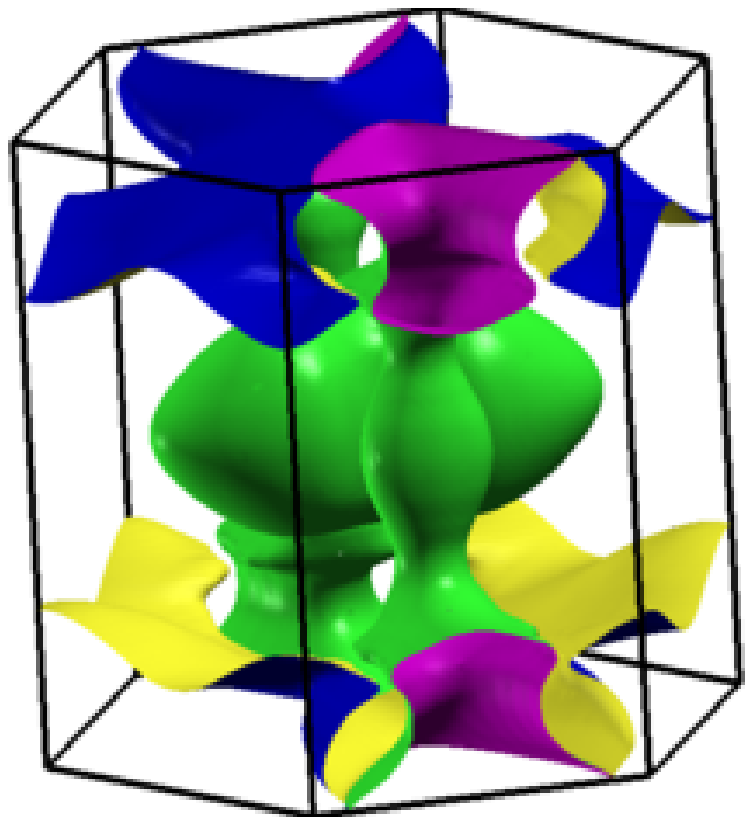}
\includegraphics[width=3cm]{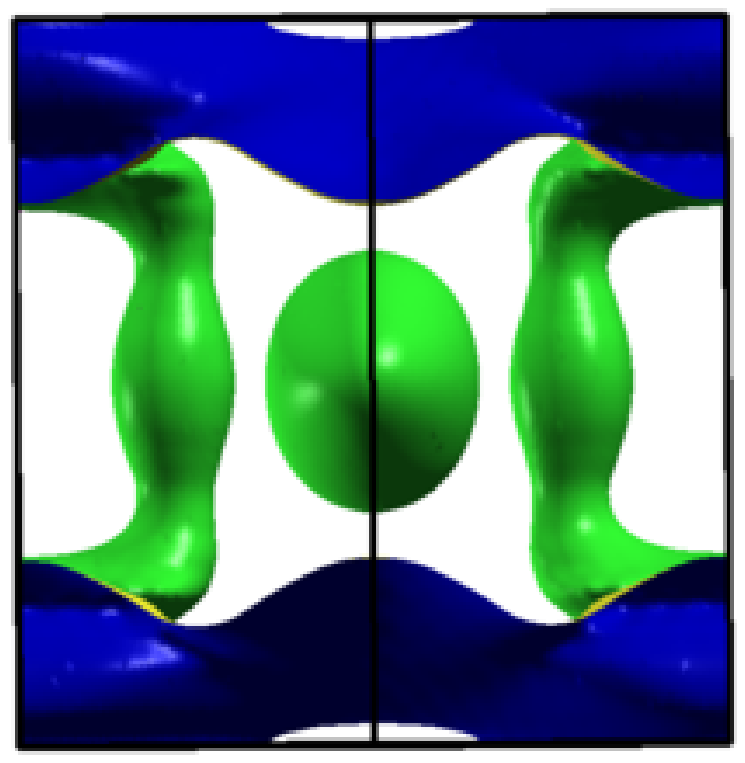}
\includegraphics[width=3cm]{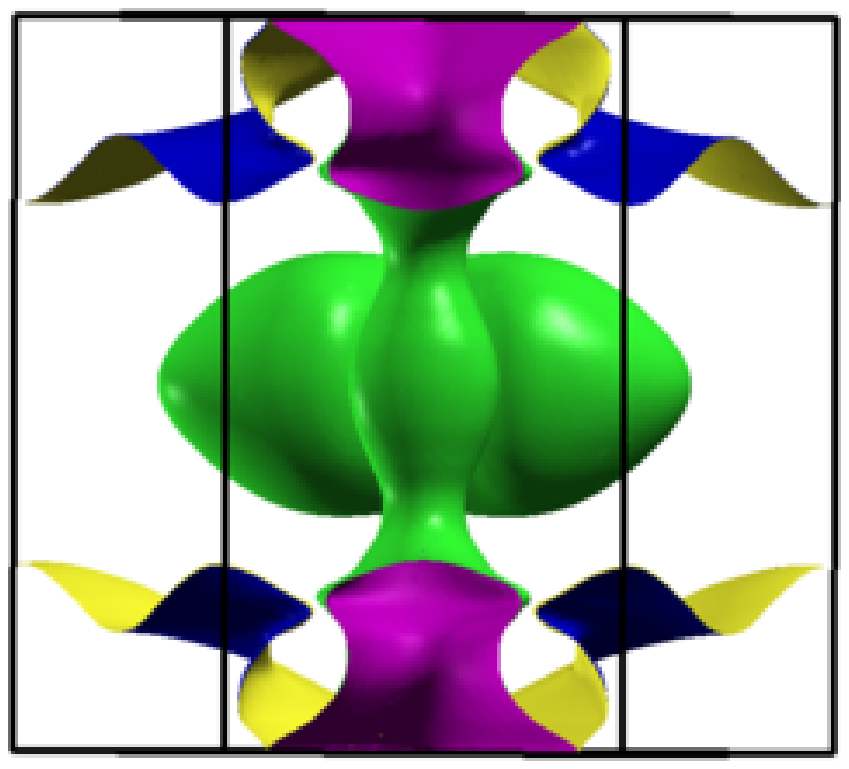}\\
\includegraphics[width=3cm]{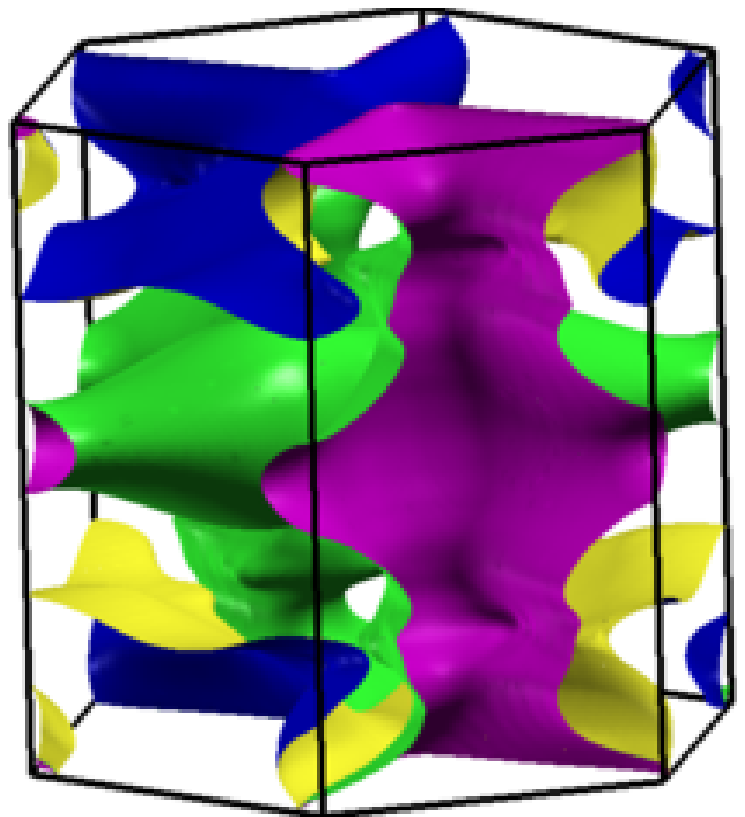}
\includegraphics[width=3cm]{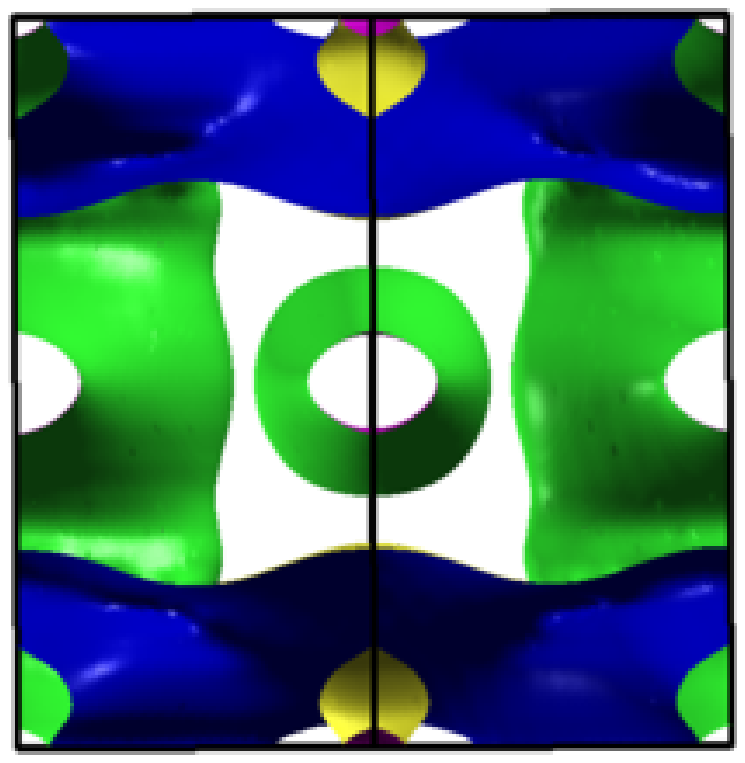}
\includegraphics[width=3cm]{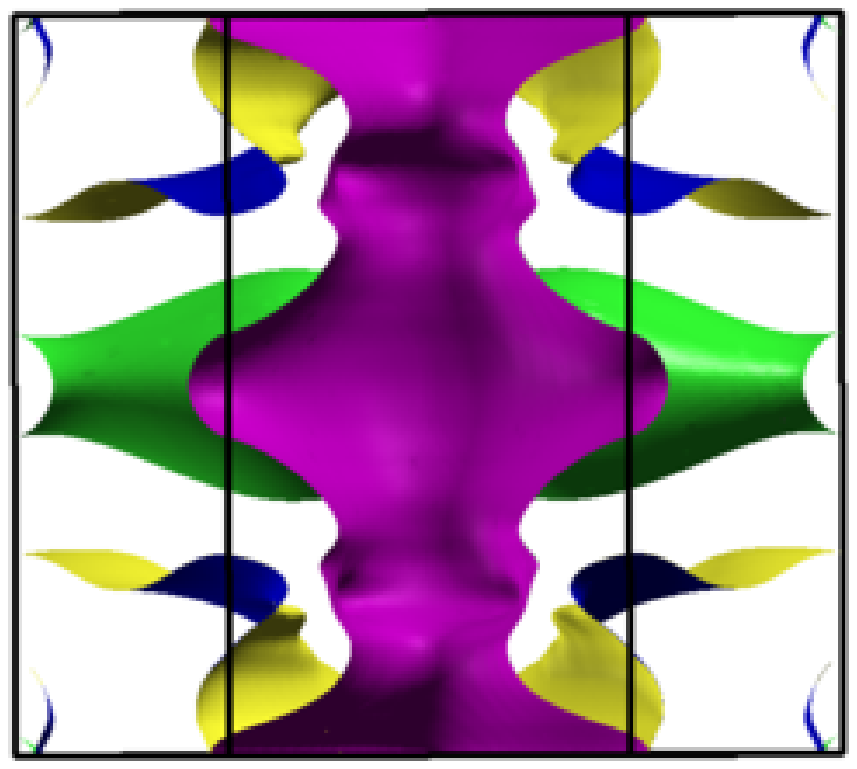}
\caption{\label{Img:LDA_FERMISURFACES} LDA
  Fermi surfaces of $\mathrm{BaVS_3}$ (top) and $\mathrm{BaVSe_3}$
  (bottom) from different perspectives.}
\end{figure}
\begin{figure}[h]
\centering
\includegraphics[width=6.5cm,clip]{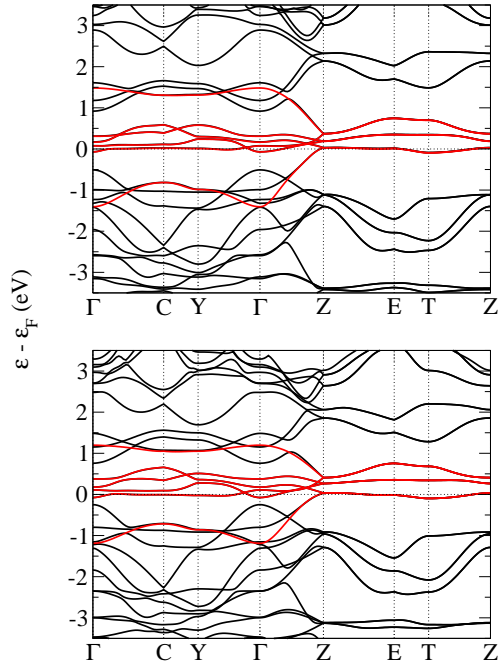}
\caption{\label{Img:BDSTRUC_LDA_WANNBANDS}
  Dispersions of the effective three-band model (red/gray), obtained by means
  of a maximally-localized Wannier-function construction, for 
  $\mathrm{BaVS_3}$~(top) and $\mathrm{BaVSe_3}$~(bottom). Black: original LDA 
  band structure.}
\end{figure}
\begin{figure}[h]
\centering
\includegraphics[width=7.5cm,clip]{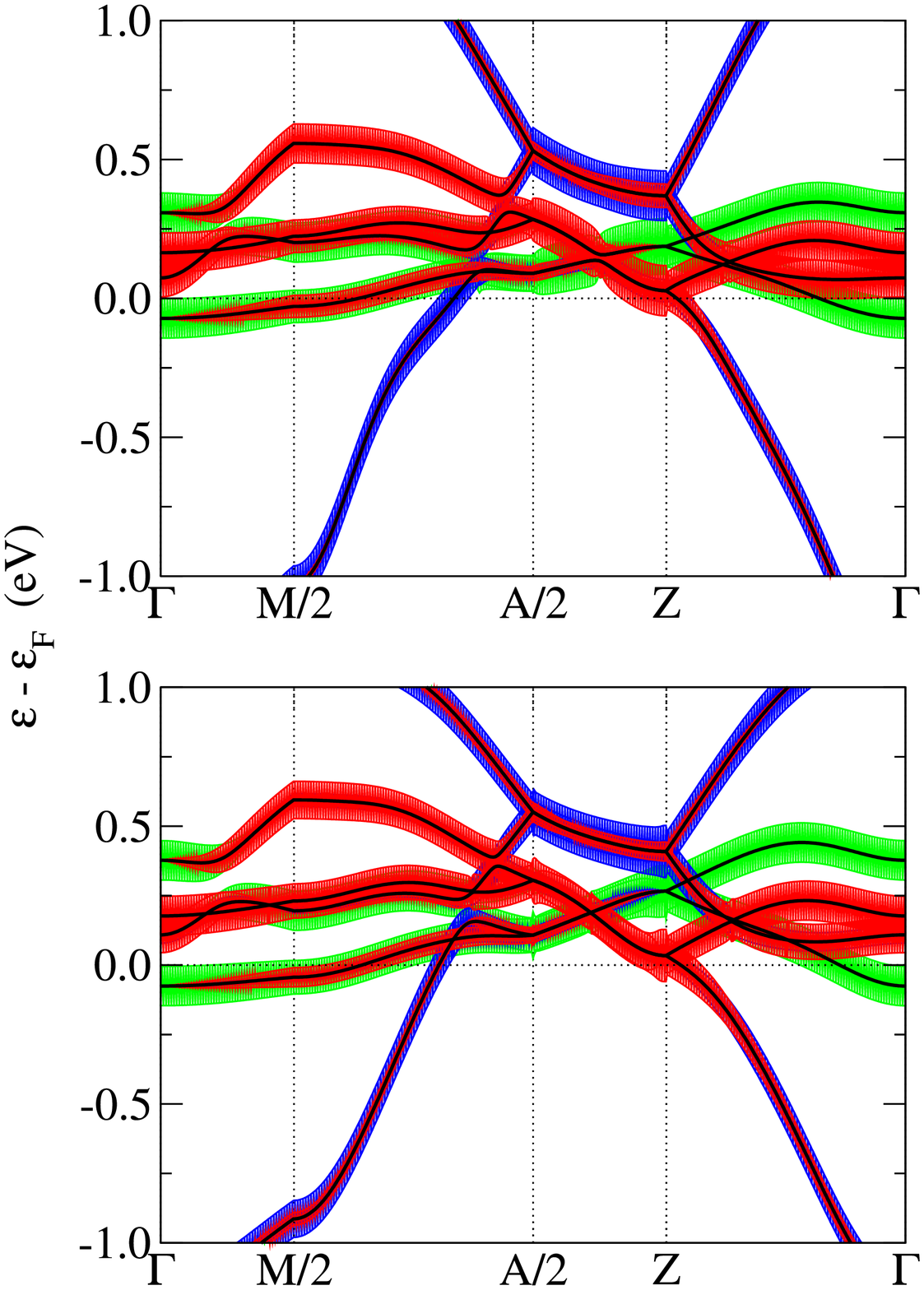}
\caption{\label{Img:WANNIER_FATBANDS_NONHIGHSYML}
  LDA $t_{2g}$ Wannier fatband plots for $\mathrm{BaVS_3}$~(top)
  and $\mathrm{BaVSe_3}$~(bottom) along a closed path in the first
  Brillouin zone, connecting $\Gamma$, '$M/2$', '$A/2$' and $Z$ 
  (see figure~\ref{Img:BDSTRUC_LDA_FATBANDS}). 
  Color coding: $A_{1g}$~(blue), $E_{g1}$~(red) and $E_{g2}$~(green).}
\end{figure}
Experimentally the nesting condition for the formation of the CDW was
detected as $2\mathbf{k}_{\mathrm
  F}$~=~$0.5\mathbf{c}^\ast$~\cite{fag03}, where $\mathbf{c}^\ast$ is
the reciprocal lattice vector along the $c$ axis.  In order to
theoretically examine this condition, it is necessary to study the
topology of the Fermi surface (FS) (see
figure~\ref{Img:LDA_FERMISURFACES}).  Therefrom the nesting condition
is not fulfilled in the pure LDA result in both materials. However
somewhat surprisingly, the quasi-1D sheet of the selenide is closer to
fulfilling the experimental nesting condition than the corresponding
sulfide sheet in LDA. In contrast, the pillar-like structures are much
more enhanced in the selenide. Hence differences in the $t_{2g}$
occupations between the two compounds may be expected.
Figure~\ref{Img:BDSTRUC_LDA_WANNBANDS} shows the respective three-band
dispersion of the correlated subspace obtained from a
maximally-localized Wannier construction~\cite{mar97} for strongly
hybridized bands~\cite{sou01}.  It should be mentioned that this
construction yields for the selenide a larger spread of the Wannier
functions (i.e. in the range of $23\,\mathrm{au}^2$ for the
$E_g$-like ones in $\mathrm{BaVSe_3}$ compared to about
$18\,\mathrm{au}^2$ for the same ones in $\mathrm{BaVS_3}$). This can
be attributed to a stronger hybridization of the $t_{2g}$ system with
the Se($4p$) electrons. Note that for the further use of this Wannier
representation we rotated the original maximally-localized Wannier
Hamiltonian into the crystal-field basis, i.e., the basis where the
on-site part thereof takes on a diagonal form.

Concerning the susceptibility towards the CDW state it is important
and evident to note that the nesting vector mentioned above is not
necessarily located along the $\Gamma$-$Z$ high-symmetry
line~\cite{lec06,lec07}.  In order to take this fact into account, the
low-energy band structure is computed along a Brillouin zone (BZ) path
containing a line parallel to the $\Gamma$-$Z$ line, shown in
figure~\ref{Img:WANNIER_FATBANDS_NONHIGHSYML}.  The points '$M/2$' and
'$A/2$' are located halfway between the $\Gamma$- or $Z$-point,
respectively, and the edge of the first BZ (see
figure~\ref{Img:BDSTRUC_LDA_FATBANDS}). From the displayed fatbands it
can be seen that the hybridization between $A_{1g}$ and $E_{g1}$ on
the resulting bands along $M/2$-$A/2$ as well as along $\Gamma$-$Z$ is
significantly larger in the selenide than in the sulfide. Thus a clear
notion of exclusive '$A_{1g}$ bands' or '$E_{g1}$ bands' becomes even
more difficult for BaVSe$_3$. This fact also impedes the assignment of
the distinct FS sheets to a single definite orbital character.

\subsection{Inclusion of correlation effects}
\begin{figure}[b]
\centering
\includegraphics[width=8.5cm,clip]{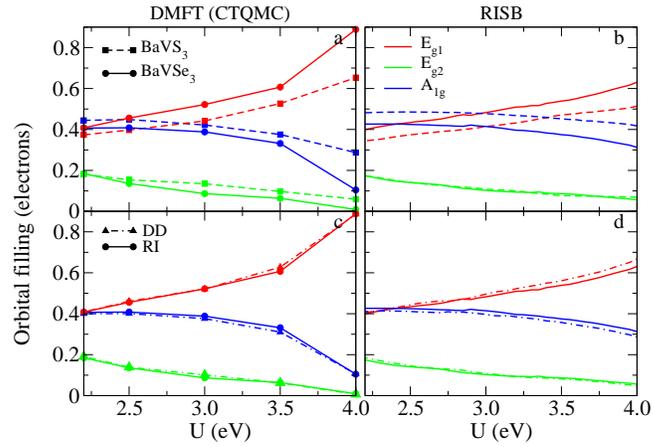}
\caption{\label{Img:U_SCAN_J07_RI_DDANDRI} (a),~(b):
  Effective $t_{2g}$ fillings for $\mathrm{BaVS_3}$ (dashed/squares) and
  $\mathrm{BaVSe_3}$ (solid/circles) using the KS-Wannier Hamiltonians with
  increasing $U$. (c),~(d): Fillings for $\mathrm{BaVSe_3}$ using only
  density-density terms (dash-dotted/triangles) and including
  spin-flip as well as pair-hopping terms (solid/circles). (a),~(c):
  CTQMC solution of DMFT and (b),~(d): RISB solution.}
\end{figure}
Besides the apparent metal-to-insulator transition in
$\mathrm{BaVS_3}$, already the complex magnetic behavior in both
systems (with different magnetic order at low temperature) shows
clearly the importance of electronic correlations in these vanadium
chalcogenides. Hence a deeper comparison of $\mathrm{BaVS_3}$ and
$\mathrm{BaVSe_3}$ has to include explicit many-body physics, which we
approached with the LDA+DMFT~(CTQMC) method as well as the LDA+RISB
technique. This twofold investigation was elaborated in order to
evaluate the reliability of the latter method compared to the more
complete CTQMC impurity solution within DMFT.  In the following, all
the presented CTQMC impurity computations were performed at $T$~=~116K
($\beta$~=~100~eV$^{-1}$) and including spin-flip and pair-hopping
terms in the interacting hamiltionian (if not stated otherwise).

In~\cite{lec05}, a chosen value of $U$~=~3.5 eV led to a proper
theoretical description of BaVS$_3$. In this work we also do not compute
the interaction parameters from first principles, but again take a
practical point of view in order to account for the key correlation
effects. As noted in the previous section, the corresponding Hubbard
$U$ for BaVSe$_3$ is expected to be smaller because of the increased
screening capabilities of the Se($4p$) electrons compared to the
S($3p$) ones. In order to compare $\mathrm{BaVS_3}$ and
$\mathrm{BaVSe_3}$ in this respect, the calculated orbital-resolved
fillings are shown in figure~\ref{Img:U_SCAN_J07_RI_DDANDRI} for
varying values of $U$ and fixed Hund's exchange $J$~=~0.7 eV. First note
the very good agreement between the DMFT~(CTQMC) and the RISB treatment
of the correlation effects. Both methods clearly show the
correlation-induced charge transfer mainly between the $A_{1g}$ and
$E_{g1}$ orbitals~\cite{lec05,lec07}. The actual numbers surely differ
somewhat between both approaches due to the neglect of the explicit
quantum fluctuations within RISB, especially concerning the frequency
dependence of the high-energy excitations. But also the trend in the
(minor) differences between density-density only terms and the
inclusion of more general terms in the interacting Hamiltonian is
still well reproduced by the simplified RISB technique. Concerning the
physics, it is visible that the $A_{1g}$ filling is generally reduced
in the selenide compared to the sulfide, which was already expected
from the Fermi-surface discussion. This result is also apparent from
table~\ref{Tab:FILLINGS_DMFT_DENSITYDENSITY}, where the occupation
numbers are displayed for selected interaction strengths. In this
context, a smaller Hubbard interaction $U$~=~2.5 eV for BaVSe$_3$ seems
reasonable in the end. However, for direct comparisons we also include
the case $U$~=~3.5 eV in the computations. Note that $J$ is usually not
strongly affected by the crystal environment and to a good
approximation can remain constant. Hence, if not stated differently, we
generally fixed the Hund's coupling to the value $J$~=~0.7 eV used in
previous studies~\cite{lec07} and which is in line with what was utilized 
in similar investigations for vanadium oxides~\cite{sol96,ole05,bie05}.
Note that a suitable energetical range for $J$ proved to be sufficient to 
describe the key low-energy physics of BaVS$_3$~\cite{lec05}. 

In summary, the system could be understood as an effective
two-band system, since the effective $E_{g2}$ orbital is almost empty,
with an even larger correlation-induced charge transfer from $A_{1g}$
to $E_{g1}$ in the selenide than in the sulfide. It can further be
seen that the spin-flip and pair-hopping terms have only a small
influence on the occupation numbers, resulting qualitatively in a
slightly smaller orbital polarization. The latter is due to the fact that
those terms lead to a further reduction of the integrated effective $U$
compared to the density-density only case. Moreover since the present
systems mainly reside in the local single-particle sector, the influence
of spin-flip and pair-hopping matrix elements on the charge fluctuations 
remains minor.

\begin{table}
\caption{\label{Tab:FILLINGS_DMFT_DENSITYDENSITY} Orbital-resolved
  fillings from LDA+DMFT~(CTQMC). Vanishing Hubbard $U$ marks the LDA result.
  'DD': only density-density interactions, 'RI': additional inclusion
  of spin-flip and pair-hopping interactions.}
\begin{indented}
\item[] \begin{tabular}{lllllll}
\br
& $U$ (eV) & $\hat H_{\mathrm{int}}$ & $A_{1g}$ & $E_{g1}$ & $E_{g2}$ \\
\mr
$\mathrm{BaVS}_3$
& 0.0 &    & 0.58 & 0.31 & 0.11\\
& 3.5 & DD & 0.38 & 0.54 & 0.08\\
& 3.5 & RI & 0.37 & 0.53 & 0.10\\
\mr
$\mathrm{BaVSe}_3$
& 0.0 &    & 0.49 & 0.40 & 0.11\\
& 2.5 & DD & 0.40 & 0.46 & 0.14\\
& 2.5 & RI & 0.41 & 0.45 & 0.14\\
& 3.5 & DD & 0.31 & 0.63 & 0.06\\
& 3.5 & RI & 0.33 & 0.61 & 0.06\\
\br
\end{tabular}
\end{indented}
\end{table}
\begin{figure}
\centering
\includegraphics[width=7.5cm,clip]{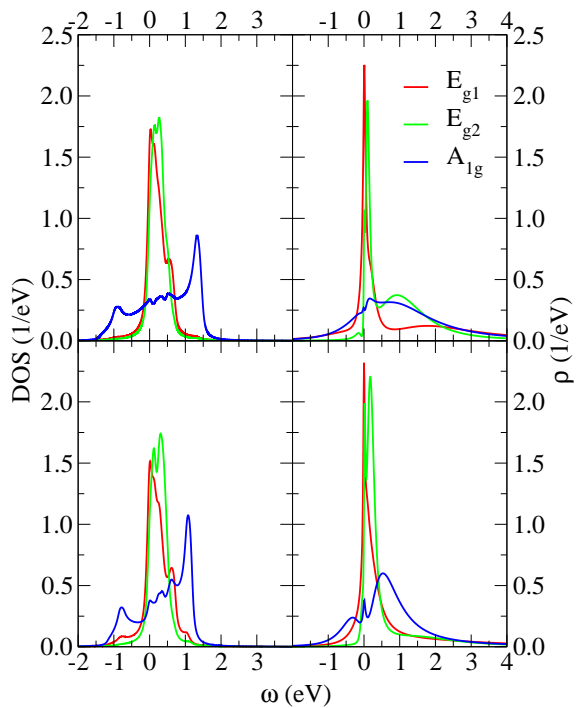}
\caption{\label{Img:BOTH_WANNIERDOS_SPECFUNC}
  Comparision of the LDA Wannier DOS (left) and local spectral functions
  from LDA+DMFT (right) for $\mathrm{BaVS_3}$ with $U$~=~3.5 eV
  (top) and $\mathrm{BaVSe_3}$ with $U$~=~2.5 eV (bottom).}
\end{figure}
Figure~\ref{Img:BOTH_WANNIERDOS_SPECFUNC} shows the density of states
derived from the original Wannier functions compared to the local
spectral functions from DMFT for both systems. On the LDA level, the
plot shows the reduced bandwidth of the selenide compound, with the
orbital resolved effective widths within the $t_{2g}$ manifold
becoming more similar. Furthermore, the overall stronger hybridization
(note the use of the crystal-field basis) between $A_{1g}$ and
$E_{g1}$ in BaVSe$_3$ is clearly visible. This also leads to a more
pronounced $A_{1g}$ quasiparticle peak right at the Fermi level in the
DMFT local spectral function.

\begin{figure}
\centering
\includegraphics[width=2.5cm]{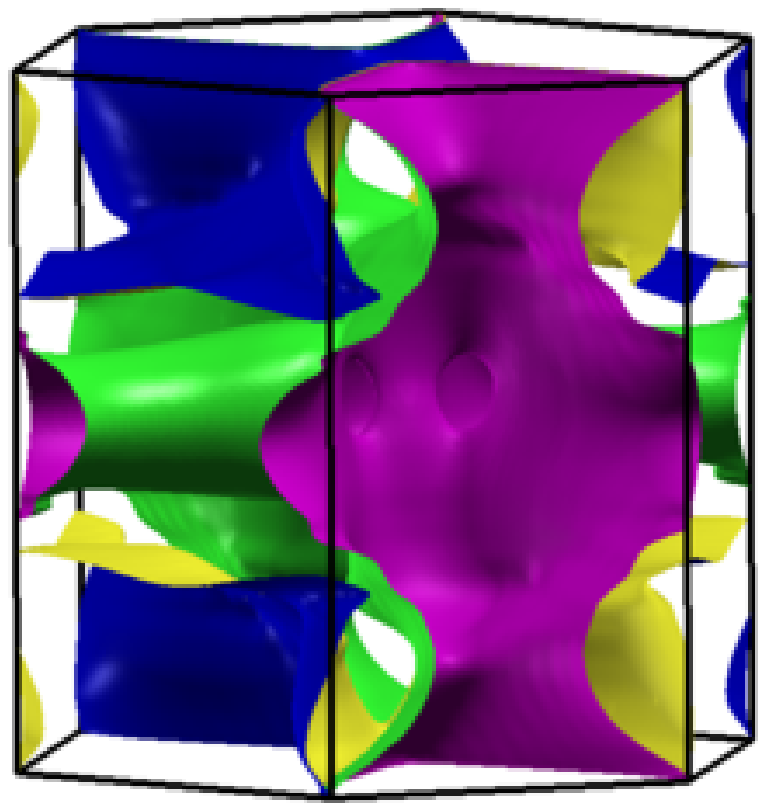}
\includegraphics[width=2.5cm]{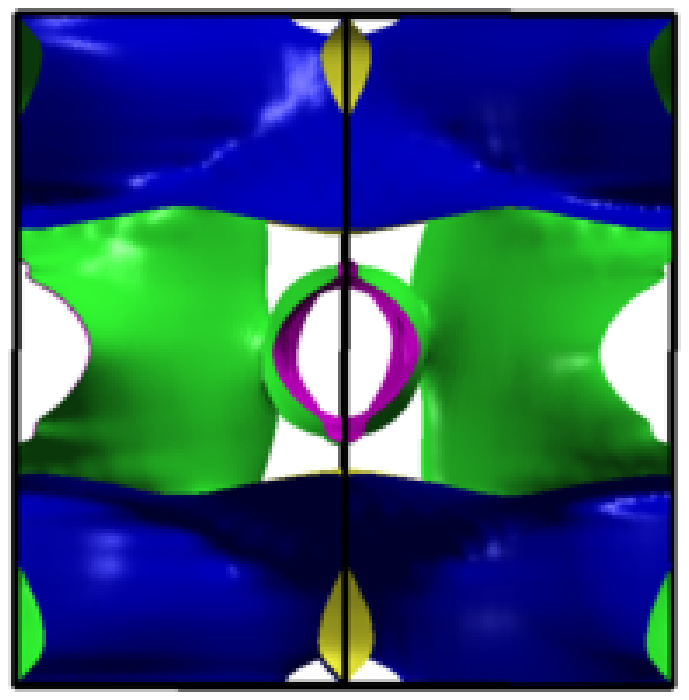}
\includegraphics[width=2.5cm]{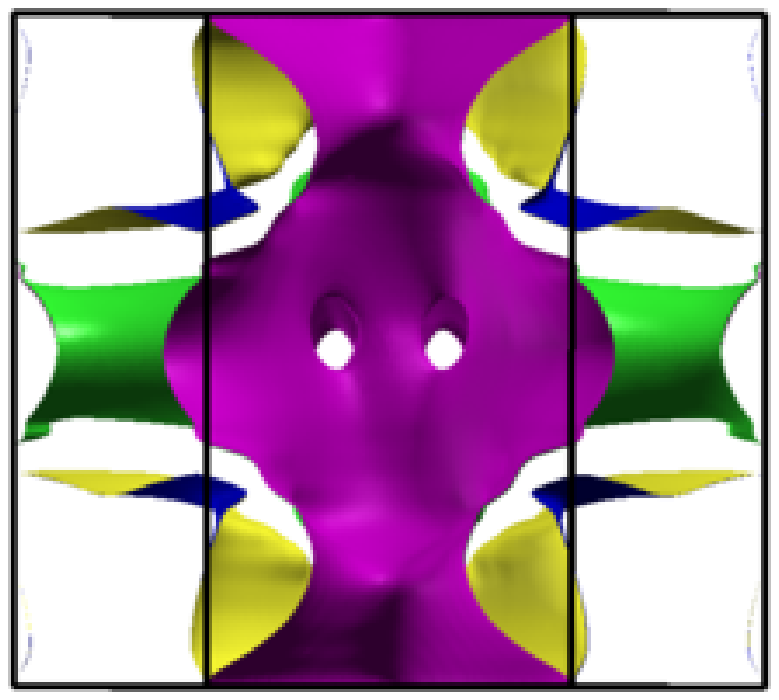}\\
\includegraphics[width=2.5cm]{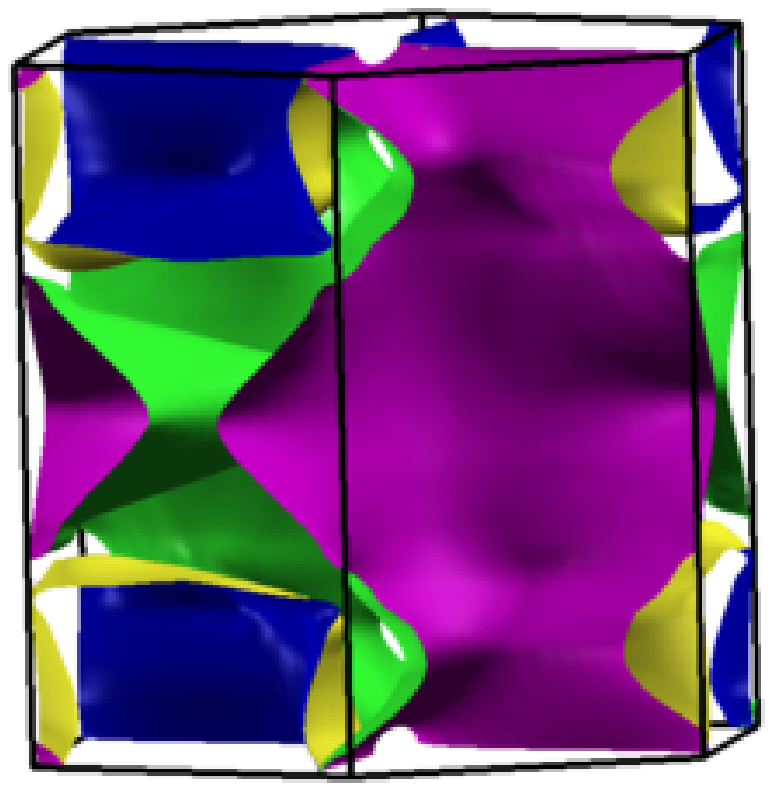}
\includegraphics[width=2.5cm]{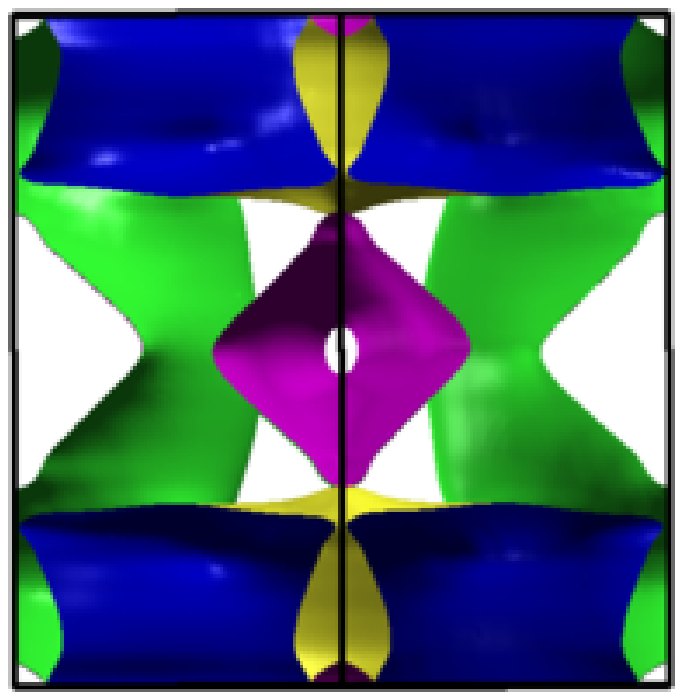}
\includegraphics[width=2.5cm]{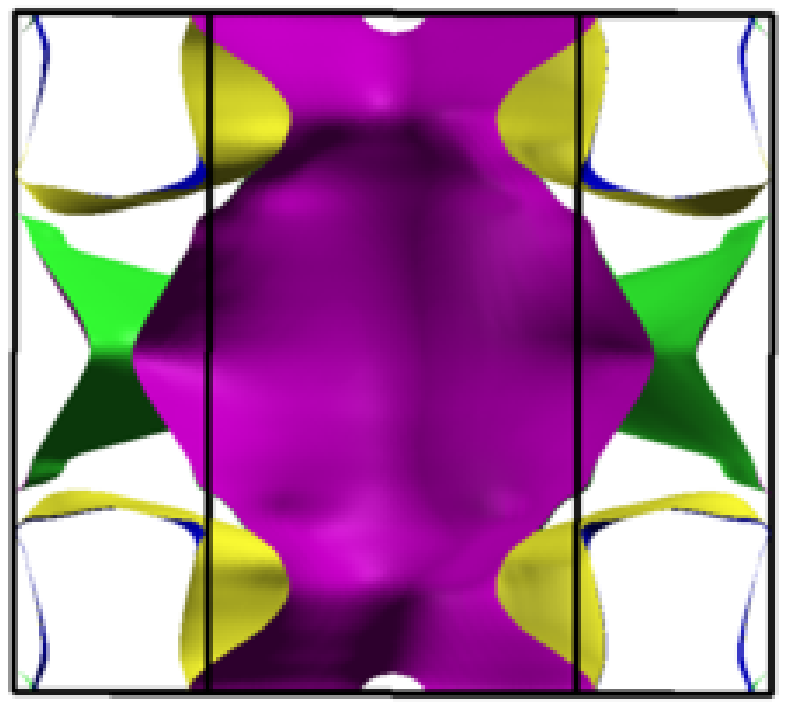}\\
\includegraphics[width=2.5cm]{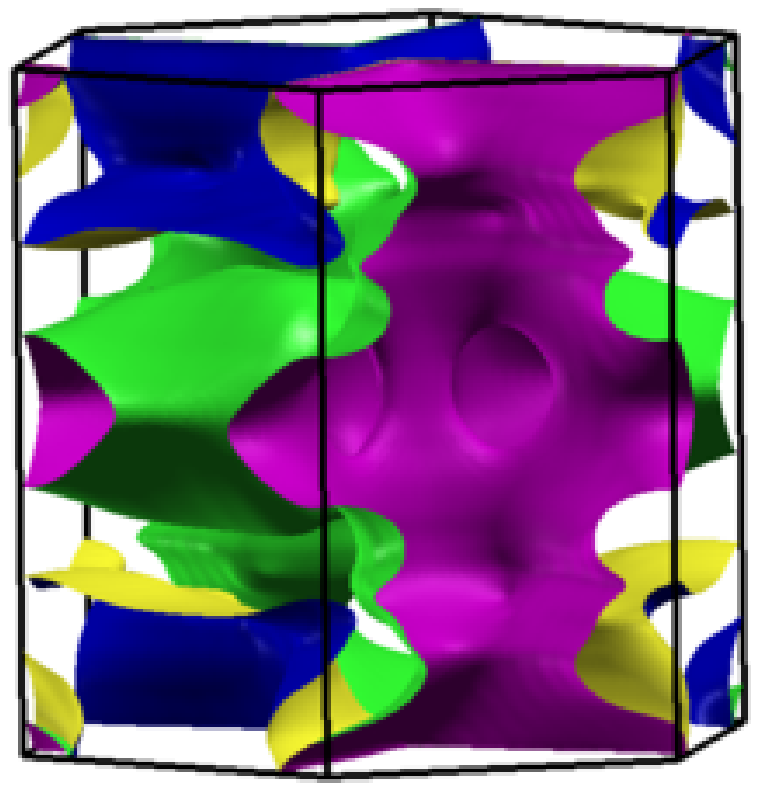}
\includegraphics[width=2.5cm]{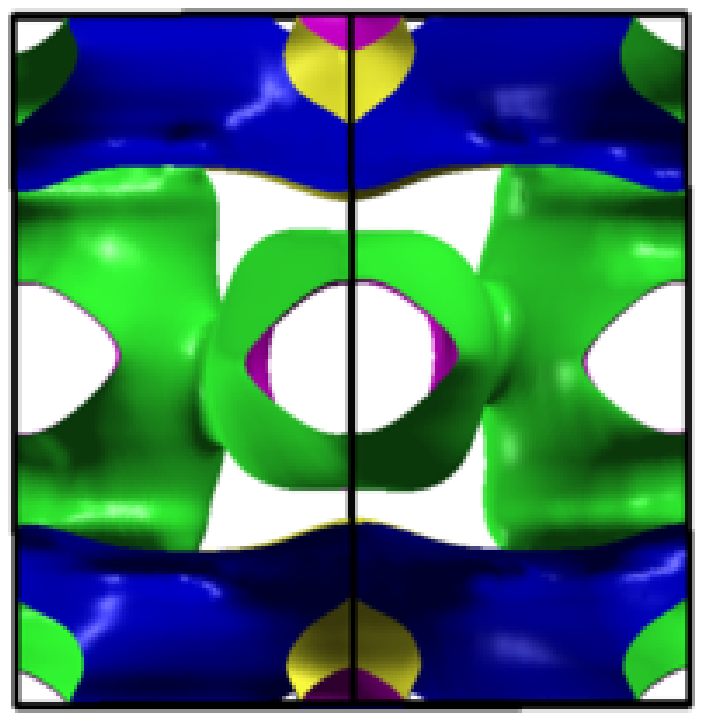}
\includegraphics[width=2.5cm]{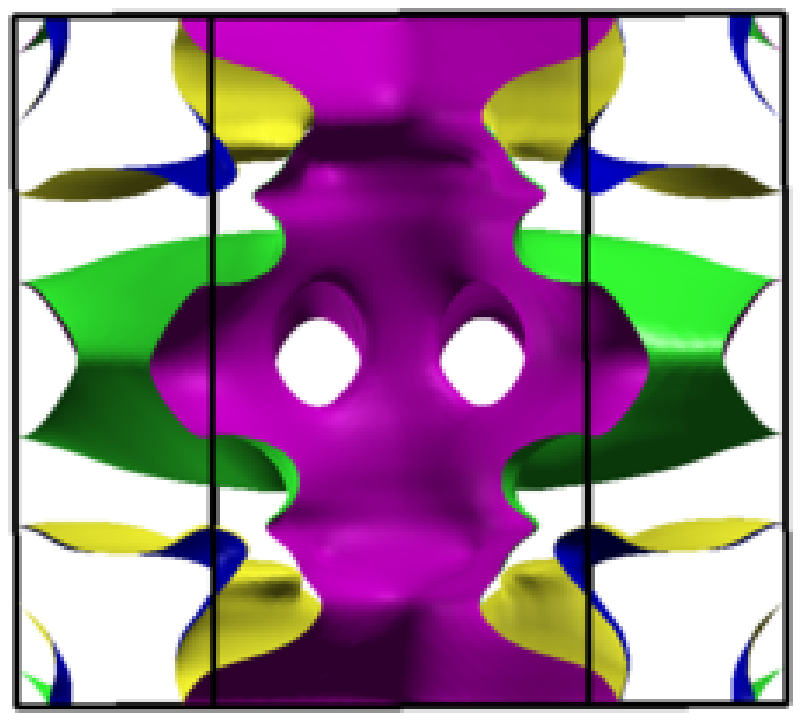}
\caption{\label{Img:DMFT_FERMISURFACES} Quasiparticle 
  Fermi surfaces of $\mathrm{BaVS_3}$ with
  $U$=3.5 eV (top) and $\mathrm{BaVSe_3}$ with
  $U$=3.5 eV (middle) and $U$=2.5 eV (bottom).}
\end{figure}
\begin{figure}[t]
\centering
\includegraphics[width=6.5cm,clip]{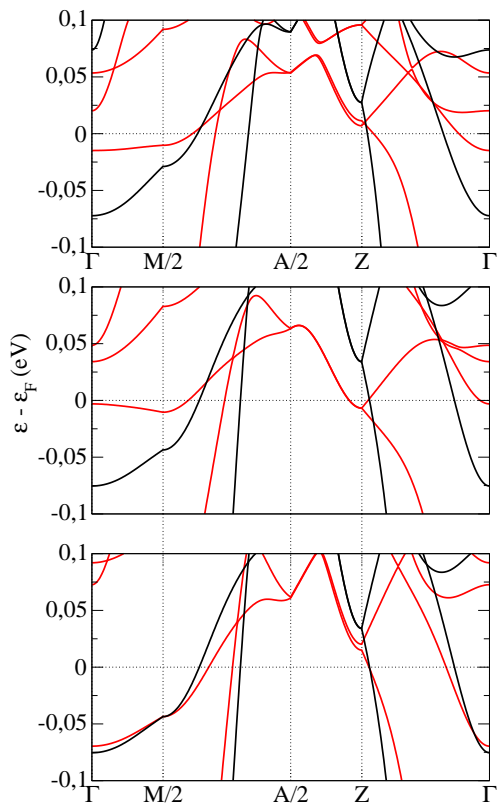}
\caption{\label{Img:BAV_BANDS_DMFT_NONHIGHSYM}
  Effective $t_{2g}$ Wannier band structure with (red/gray) and
  without (black/dark) renormalization from LDA+DMFT along the closed
  path including '$M/2$' and '$A/2$' (see
  figure~\ref{Img:BDSTRUC_LDA_FATBANDS}) for BaVS$_3$ (top) and BaVSe$_3$
  with $U$=3.5~eV (middle) and $U$=2.5~eV (bottom).}
\end{figure}

As explained in~\cite{lec05,lec07}, the importance of explicit
many-body effects in deforming the LDA Fermi surface appears as a
crucial ingredient for the onset of the CDW in BaVS$_3$. Therefore,
figure~\ref{Img:DMFT_FERMISURFACES} shows the DMFT~(CTQMC)
quasiparticle Fermi surfaces for both chalcogenides.
Correspondingly, figure~\ref{Img:BAV_BANDS_DMFT_NONHIGHSYM} displays
the quasiparticle dispersion for the closed path in the first BZ
connecting the points '$M/2$' and '$A/2$' (see
figure~\ref{Img:BDSTRUC_LDA_FATBANDS}), which mark the relevant
distance for the possible CDW nesting. The plots render it obvious
that no definite information about the absence of the CDW in BaVSe$_3$
can be retrieved from this viewpoint. Since qualitatively the same 
charge-transfer mechanism applies to the selenide compound, it may be observed 
that the FS deformations also go in the same direction. This
means that within the LDA+DMFT accuracy also for smaller Hubbard $U$
the experimental nesting condition could in principle be realized in BaVSe$_3$. 
Note however that in the selenide for $U$=2.5~eV the respective quasi-1D
sheets are still not strongly flatted.

In order to get a quantitative comparative understanding of the degree
of quasi-one-dimensionality of the two materials, we investigated the
theoretical order of magnitude of the anisotropy in the DC
conductivities in LDA as well as in the strongly correlated case. To
this end, table~\ref{Tab:FERMIVELOCITIES} summarizes the relation of
the spatial components of the averaged Fermi velocities squared
($\langle \mathbf v_F^2 \rangle = \langle v_x^2 \rangle + \langle
v_y^2 \rangle + \langle v_z^2 \rangle$), whereas $y$ denotes the
spatial direction in which the zig-zag distortion of the vanadium
chains can be observed. Note that the LDA values obtained for the
sulfide are in perfect agreement with the earlier result from
Mattheiss~\cite{mat95}. Surprisingly, the selenide shows an even
larger formal DC anisotropy in LDA compared to the sulfide
compound. In contrast, it is also obvious that within the many-body
approach of LDA+DMFT the renormalization of the anisotropy, especially
in the dominant $\langle v_z^2 \rangle / \langle v_x^2 \rangle$
channel leads to a balanced value for BaVSe$_3$ with reasonable
$U$=2.5 eV. In contrast, the quasiparticles become indeed
significantly heavier in the $c$ direction for BaVS$_3$ with
$U$~=~3.5~eV. It is therefore worthwhile and interesting to note that the
sulfide compound indeed displays a clear signature of DC anisotropy in
the correlated case, whereas such a clear character is appearing in
the according BaVSe$_3$ data only at improper large $U$. It shall be
remarked that the intraorbital quasiparticle weight for BaVS$_3$
(BaVSe$_3$) amounts to $Z$~$\sim$~0.5 (0.6) with minor differences
within the $t_{2g}$ manifold. Note that, although we work in the
crystal-field basis with a diagonal onsite KS Hamiltonian for each of
the two V atoms in the unit cell, the interacting terms together with
the hybridization between these atoms lead to additional offdiagonal
(interorbital) self-energy terms between $A_{1g}$ and $E_{g1}$ for the
present crystal symmetry. Both many-body approaches, i.e. DMFT~(CTQMC)
and RISB, are capable of revealing and handling this effect. Thus a
diagonal Green's function approach is invalid in the present case.

%

\begin{table}[t]
\caption{\label{Tab:FERMIVELOCITIES} Relation of the spatial
  components of squared Fermi velocities in $\mathrm{BaVS_3}$ and
  $\mathrm{BaVSe_3}$ on the LDA level and within LDA+DMFT~(CTQMC) for a
  given $U$ (in eV).}
\begin{indented}
\item[]\begin{tabular}{lrr|rrr}
\br
& \multicolumn{2}{c|}{$\mathrm{BaVS_3}$} & 
\multicolumn{3}{c}{$\mathrm{BaVSe_3}$}\\
Ratio      & LDA & $U$=3.5 & LDA & $U$=2.5 & $U$=3.5 \\
\mr
$\langle v_z^2 \rangle / \langle v_x^2 \rangle$ & 3.7 & 2.7 & 4.8  & 5.2  &  3.0 \\
$\langle v_z^2 \rangle / \langle v_y^2 \rangle$ & 8.3 & 9.9 & 10.7 & 9.9 & 10.2\\
\br
\end{tabular}
\end{indented}
\end{table}

\begin{figure}[t]
\centering
\includegraphics[width=8.5cm,clip]{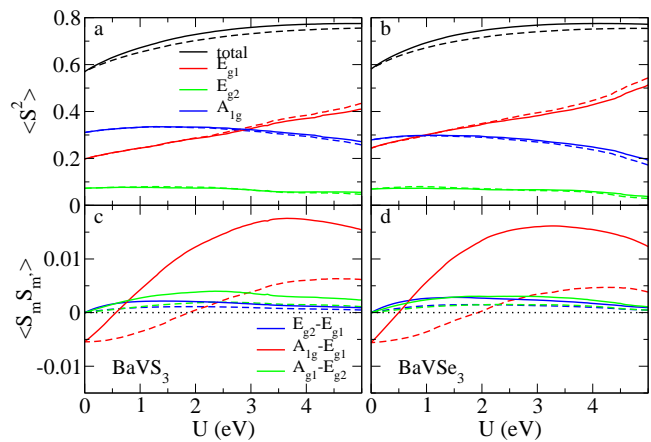}
\caption{\label{Img:SPIN_COMPARE} On-site spin correlations for
 BaVS$_3$ (left) and BaVSe$_3$ (right). Dashed lines: density-density only 
 interactions, solid lines: including also spin-flip and pair-hopping terms. 
 In each $U$ scan a ratio $U$/$J$=5 was chosen.
 (a-b): Total and orbital-resolved diagonal spin-spin expectation value.
 (c-d): Off diagonal spin-spin correlation between the effective $t_{2g}$ 
 orbitals.}
\end{figure}

Finally, since the spin degree of freedom plays an additional vital role in 
these chalcogenide compounds, figure~\ref{Img:SPIN_COMPARE} exhibits the 
respective on-site spin-correlation functions from the LDA+RISB computations 
for constant ratio $U/J$=5. Concerning the diagonal spin-spin expectation 
value, as expected, the magnitudes evolve according to the nominal orbital 
occupations. Hence $\langle S^2\rangle$ for the effective $E_{g1}$ orbital 
dominates over the corresponding value for the effective $A_{1g}$ orbital in 
the strong-interaction limit. In this respect, again quantitatively BaVSe$_3$ 
exceeds BaVS$_3$ due to the increased $E_{g1}$ occupation. The total 
$\langle S^2\rangle$ increases with larger $U$ due to the growing electron 
localization. Note that in the present problem the magnitude of this diagonal 
quantity can exceed the atomic-limit value of a single electron, i.e., 
$\langle S^2\rangle_{\rm at}$=$\frac{1}{2}(\frac{1}{2}+1)$=$\frac{3}{4}$, 
albeit we have nominally one electron in the $3d$ shell of the chalcogenides. 
This is due to the fact that at some point a strong Hund's coupling may 
tolerate an occupation of two electrons in the local three-orbital manifold of
a metal. Since two electrons with parallel spin alignment form an $S$=1 
system, the resulting $\langle S^2\rangle$ is bigger than the sum of two 
individual single-electron occupied sites. The interorbital spin-spin 
correlations again prove the dominant $A_{1g}$-$E_{g1}$ hybridization, which 
also leads to a significant spin coupling. However, it may be observed that
in the minimal modeling for both systems the first Hund's rule is violated  
close to the $U$=0, i.e., the offdiagonal 
$\langle S_{A_{1g}}S_{E_{g1}}\rangle$ becomes smaller than zero in this 
regime. 
\begin{figure}[t]
\centering
\includegraphics[width=8.5cm,clip]{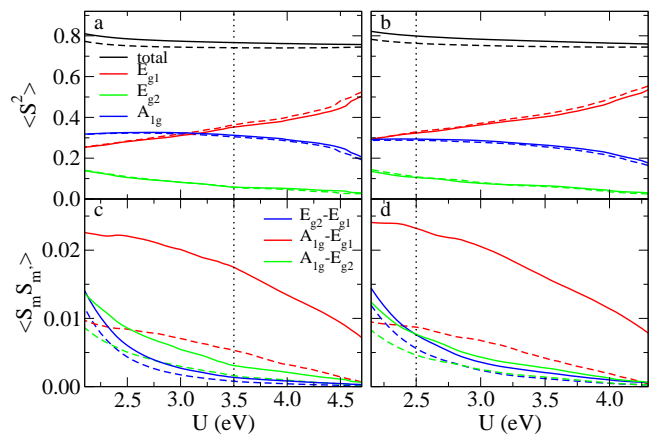}
\caption{\label{Img:SPIN_COMPARE_J} Same as 
figure~\ref{Img:SPIN_COMPARE}, but with fixed Hund's coupling $J$=0.7 eV. The
vertical dotted lines indicate the expected suitable choice for the Hubbard $U$
of the respective system.}
\end{figure}

It also has to be noted that these interorbital spin-spin correlations
are rather sensitive to the degree of approximation of the rotational
invariance in the interaction. While the diagonal spin-spin
expectation value is not strongly affected by the inclusion of
spin-flip and pair-hopping terms in the local Hamiltonian, the
offdiagonal $\langle S_m S_{m'}\rangle$ is strongly enhanced in this
case by a factor of two to three. Thus, whereas integrated one-particle
quantities like occupation numbers and local spins do not strongly
suffer from density-density only descriptions, the two-particle
functions, i.e. susceptibilities, may be substantially
different. Figure~\ref{Img:SPIN_COMPARE_J} displays the same spin-spin
functions, but for the (more realistic) case of a constant $J$=0.7
eV. Of course, the strongest response may be observed for smaller $U$
since then the Hund's coupling is most effective. Note again the
differences for the density-density only interactions, leading to an
even qualitatively different behavior for small $U$, where $\langle
S_{A_{1g}}S_{E_{g1}}\rangle$ is eventually dominated by the $\langle
S_{E_{g2}}S_{E_{g1}}\rangle$. In the more sound close to rotationally
invariant description however, $\langle S_{A_{1g}}S_{E_{g1}}\rangle$
is always clearly strongest. Because of the supposedly different $U/J$
ratio for BaVS$_3$ and BaVSe$_3$ we hence expect for the selenide compound
a more distinct magnetic behavior.
Note that the interorbital spin-spin correlations decay in any case for 
large $U$, contrary to what is known for half-filled systems (there  
$\langle S_m S_{m'}\rangle$ actually increases with $U$). This is due to 
the fact that the ground state for a multi-orbital problem with a 
single-electron filling corresponds at large $U$ to the localization of 
that electron within a single orbital.

\section{Discussion}
From the above calculations, it turns out that $\mathrm{BaVS_3}$ and
$\mathrm{BaVSe_3}$ are in many respects very similar materials. The
differences between them are rather subtle, but still large enough to
be resolved by the applied methods. The picture that arises thereof is
not totally unambiguous, but several hints can be given why the
formation of a charge-density wave is hindered in
$\mathrm{BaVSe_3}$. Already on an LDA level, it can be seen that
hybridization effects among the bands of the $t_{2g}$ multiplet as
well as with the corresponding S($3p$)/Se($4p$) electrons, are
significantly stronger in the selenide than in the sulfide. Hence a
manifest discrimination between dominant $A_{1g}$ and dominant
$E_{g1}$ character is far less obvious than in BaVS$_3$. Furthermore, the
strong role of the effective $A_{1g}$ orbital is weakened already on
the LDA level and the final dominance of the effective $E_{g1}$
orbital in the correlated case is even enhanced. This is altogether
underlined by the DC anisotropies for the selenide, that under the
effect of the expected correlations do not show the highlighting
renormalizations in the $c$ direction. Thus although the orbital
differences are obvious, it seems as if the $t_{2g}$ manifold in BaVSe$_3$
acts somewhat more cooperatively compared to the largely competing scenario in
BaVS$_3$.

With a surely smaller Hubbard $U$ the selenide compound should also be closer
to the LDA limit~\cite{akr08}, with a less pronounced nesting susceptibility
towards the CDW state. However as described, from a pure Fermi-surface 
discussion no clear answer to this question may be conveyed. Nevertheless,
electronic correlations are not irrelevant in BaVSe$_3$ as they underline the
dominance of the $E_{g1}$ level and the eventual importance of magnetic 
correlations. These latter are again stronger in the selenide compared to the
sulfide, at least from a local viewpoint. Further calculations of intersite
spin-spin correlations should shed more light onto this.

In the end, the present study within standard LDA+DMFT is not in the
definite position to clearly assign the different experimental facts
of these suprisingly similar compounds from a theoretical
perspective. However, although the situation is very subtle, we
believe we have shown several indications that may motivate to a
certain extent why the present compounds behave so differently at
lower temperatures. There seem to exist two main possible
instabilities in these chalcogenides, namely the CDW one and the
(ferro)magnetic ordering instability. As the driving forces towards
these broken-symmetry states compete, a tailored renormalization-group
study could reveal the finally dominating instability channel for each
compound. Hence it is expected that the minor differences revealed in
the present investigation show up more evidently within a renormalized
scaling scenario. Still it has to be appreciated that, though not fully
comprehensive, the LDA+DMFT method is nowadays on a level where it is
feasible to reveal fine details between rather similar materials.

\ack We wish to thank S.~Schuwalow and C.~Piefke for many fruitful
discussions concerning the technical aspects of the theoretical
approach.  Furthermore we are indebted to O.~Parcollet and M.~Ferrero
for providing their CTQMC implementation as well as to A.~Akrap and
V.~Ilakovac for illuminating insights into recent experimental results
obtained for these chalcogenide systems. Computations have been
performed at the Regionales Rechenzentrum (RRZ) of the Universit\"at
Hamburg as well as the Norddeutscher Verbund f\"ur Hoch- und
H\"ochsleistungsrechnen (HLRN).

\section*{References}

\bibliographystyle{iopart-num}
\bibliography{bibextra}

\end{document}